\documentclass[a4paper,superscriptaddress,twocolumn,english,prb]{revtex4}
\usepackage[latin9]{inputenc}
\usepackage{amsmath}
\usepackage[usenames,dvipsnames]{color}

\makeatletter
\@ifundefined{textcolor}{}
{%
\definecolor{BLACK}{gray}{0}
\definecolor{WHITE}{gray}{1}
\definecolor{RED}{rgb}{1,0,0}
\definecolor{GREEN}{rgb}{0,1,0}
\definecolor{BLUE}{rgb}{0,0,1}
\definecolor{CYAN}{cmyk}{1,0,0,0}
\definecolor{MAGENTA}{cmyk}{0,1,0,0}
\definecolor{YELLOW}{cmyk}{0,0,1,0}
}

\makeatother

\usepackage{babel}

\newcommand{\qenergies}{{quasi-energies}}

\newcommand{\Eq}[1]{Eq.~(\ref{#1})}
\newcommand{\Eqs}[1]{Eqs.~(\ref{#1})}
\newcommand{\vac}{|{\rm Vac}\rangle}

\begin{document}

\title{Poor man's derivation of the Bethe-Ansatz equations for the Dicke
model}
\author{Oleksandr Tsyplyatyev}
\affiliation{Department of Physics, University of Basel, Klingelbergstrasse 82,
CH-4056 Basel, Switzerland}
\author{Jan von Delft }
\affiliation{Arnold Sommerfeld Center and Center for Nano-Science, 
Ludwig-Maximilians-University, Theresienstr. 37, Munich, D-80333, Germany}
\author{Daniel Loss}

\affiliation{Department of Physics, University of Basel, Klingelbergstrasse 82,
CH-4056 Basel, Switzerland}
\begin{abstract}
 We present an elementary derivation of the exact solution (Bethe-Ansatz equations) of the Dicke model, using only commutation relations and 
 an informed  Ansatz for the structure of its eigenstates. 
\end{abstract}

\date{\today}

\maketitle
In 1954, Dicke showed that a model describing a set of two-level
systems coupled to a quantised electromagnetic mode leads to a
supperradient effect \cite{Dicke}.  Generalisations to
multicomponent systems naturally appear in various experimentally relevant
contexts\cite{Haroche,MultiTheory}.  Other generalisations involve a
spatially extended photonic field\cite{RupasovYudson}, or itinerant
two level systems\cite{Simons}, motivated by an experiment on cold
atoms in a 2D lattice coupled to an optical resonator\cite{Esslinger}.

The Bethe-Ansatz solution for the Dicke model with inhomogeneous
excitation energies was originally obtained by Gaudin \cite{Gaudin} as
a side result of solving the central spin problem. Using a variational
method that results in complex algebraic computations, he showed that
the solution of the central spin problem is equivalent to the
Bethe-Ansatz solution of the BCS problem derived by
Richardson\cite{Richardson}.  By expanding the Bethe-Ansatz equations
for the central spin model in the limit of large central spin, Gaudin
obtained corresponding equations for the Dicke model\cite{Gaudin}. Though this
procedure solves the original problem, the derivation is
computationally complex, and thus not easily extended to other,
related models.

The purpose of this paper is to provide an elementary derivation that
starts from the original Dicke model, in the hope that our simplified
treatment might pave the way toward finding similar solutions
to generalized Dicke models. We follow a method suggested by
Richardson\cite{Richardson1998} for the BCS model and presented in
Ref.~\onlinecite{BraunvonDelft,vonDelftRalph}.  This methods exploits the
observation that the structure of the exact eigenstates of the Dicke
model is similar to that of an auxiliary model, involving only bosons.
The only difference is that the eigenvalue equations
that determine the \qenergies\ characterizing 
these states become more complicated for the Dicke model: they  turn into
Gaudin's Bethe-Ansatz equations, which we derive here using
only commutation relations.

The inhomogeneous Dicke model describes a 
set of non-identical two-level systems with excitation energies $\epsilon_{j}$
and a single photon mode with frequency $\omega$, coupled with interaction strength $g$: 
\begin{equation}
 H=\omega b^{\dagger}b+\sum_{j=1}^{N}\epsilon_{j}\left(S_{j}^{z}+{\textstyle 
\frac{1}{2}}\right)+g\sum_{j}\left(S_{j}^{+}b+S_{j}^{-}b^{\dagger}\right) . 
\label{eq:H}
\end{equation}
The spin-$\frac{1}{2}$ operators satisfy $(S_j^\pm)^2 = 0$
and 
\begin{eqnarray}
\label{eq:spin-1/2}
[S_i^-, S_j^+] = - 2 S_j^z \delta_{ij} , \qquad
[S_i^z, S_j^\pm] = \pm S_j^\pm \delta_{ij} , \qquad
\end{eqnarray}
while the boson operators  satisfy $\left[b,b^\dagger \right]=1$.

Let $\vac$ be the ``vacuum'' state containing no boson excitations and
all spins down, i.e. $b \vac = S_j^- \vac = 0$.   $H$
  commutes with the operator $b^\dagger b+\sum_{j=1}^N S^z_j$, which
  counts the number of excitations relative to $\vac$. Thus, $H$-eigenstates can
  be constructed by acting on $\vac$ with (products of) linear
  combinations of $S_j^+$ and $b^\dagger$ operators, of the general
  (unnormalized) form
\begin{eqnarray}
\label{eq:defineB_nu}
B^\dagger_\nu = b^\dagger + \sum_{j = 1}^N A_{\nu j} S_j^+ \; ,
\end{eqnarray}
where the coefficients $A_{\nu j}$ are to be determined. For an
eigenstate with $n$ excitations relative to $\vac$ we thus make the
Ansatz (following \cite{BraunvonDelft,vonDelftRalph}),
\begin{eqnarray}
\label{eq:Ansatz}
|\Psi_n \rangle = P_1^n \vac \; , 
\end{eqnarray}
where we use the shorthand notation (for $n' \le n$)
\begin{eqnarray}
 \label{eq:Pproducts}
 P_{n'}^n = \prod_{\nu = n'}^n B_\nu^\dagger \;  
\end{eqnarray}
for a product of $B^\dagger$'s (for $n'>  n$, we set $P_{n'}^n = 1$).
For later use, note that such products satisfy
the  composition rule $P_{n'}^\nu P_{\nu+1}^n = P_{n'}^n$ for $n' \le \nu < n$.

We require that $H |\Psi_n \rangle = {\cal E}_n | \Psi_n \rangle$. 
Commuting $H$ past $P_1^n$ to the right and using
$H \vac = 0$, we obtain 
\begin{eqnarray}
 \label{eq:Hn=En}
 \left( {\cal E}_n P_1^n - [H, P_1^n ] \right) \vac = 0 \; .
\end{eqnarray}
Using the general operator identity
\begin{eqnarray}
 \label{eq:commuteXpastP}
 [X,P_{n'}^n] = \sum_{\nu = n'}^n P_{n'}^{\nu -1} [X, B^\dagger_\nu ] P^n_{\nu+1} \; , 
\end{eqnarray}
\Eq{eq:Hn=En} can be written as 
\begin{eqnarray}
 \label{eq:Hn=Enexplicit}
 \left( {\cal E}_n P_1^n - \sum_{\nu=1}^n P_1^{\nu-1} [H, B_\nu^\dagger] P_{\nu+1}^n
\right) \vac = 0 \; .
\end{eqnarray}
The requisite commutator is given by
\begin{eqnarray}
\left[H,B_{\nu}^{\dagger}\right]  =  
\sum_{j=1}^N (A_{\nu j} \epsilon_j + g) S_j^+ + 
(\omega - 2 g X_\nu)b^\dagger ,
\label{eq:HcommutatorBnu}
\end{eqnarray}
where 
$ X_\nu = \sum_{j=1}^N A_{\nu j} S_j^z$. 
By making the choice
\begin{eqnarray}
 \label{eq:chooseAnuj}
 A_{\nu j } =  \frac{g}{E_\nu - \epsilon_j } \; ,
\end{eqnarray}
where the parameters $E_\nu$ will be called \qenergies, \Eq{eq:HcommutatorBnu}
can be brought into the simplified form
\begin{eqnarray}
 \label{eq:simplifiedcommutator}
 \left[H,B_{\nu}^{\dagger}\right]  =   E_{\nu}B_{\nu}^{\dagger} + (\omega - E_\nu - 2 g X_\nu) b^\dagger \; .
\end{eqnarray}
Inserting this into \Eq{eq:Hn=Enexplicit} and identifying
the eigenergy with the sum on \qenergies, 
${\cal E}_n = \sum_{\nu = 1}^n E_\nu$, yields
\begin{eqnarray}
 \label{eq:condition1}
 \sum_{\nu=1}^n P_1^{\nu-1} (\omega - E_\nu - 2 g X_\nu) P_{\nu+1}^n
 b^\dagger \vac = 0 \; .
\end{eqnarray}
To make sense of this condition consider, for a moment, an auxiliary,
purely bosonic model, obtained from the Dicke Hamiltonian (\ref{eq:H})
by replacing $S_j^+$, $S_j^-$ and $(S_j^z + \frac{1}{2})$ by
$b_j^\dagger$, $b_j$ and $b_j^\dagger b_j$, respectively, with $[b_i,
b_j^\dagger ] = \delta_{ij}$. Repeating the above analysis yields only
one change: since $[b_j, b_j^\dagger]$ gives 1 instead of $[S_j^-,
S_j^+]$ giving $-2S_j^z $, the \emph{operator} $X_\nu$ in
\Eq{eq:HcommutatorBnu} is replaced by the \emph{c-number} $x_\nu =
-\frac{1}{2} \sum_{j=1}^N A_{\nu j}$. Thus \Eq{eq:condition1} can be
satisfied by requiring that $ \omega - E_\nu -2 g x_\nu = 0$ for all
$\nu$. Via \Eq{eq:chooseAnuj} this implies
$ \omega - E_\nu + \sum_{j = 1}^N g^2/(E_\nu - \epsilon_j)  = 0$, 
which determines the $E_\nu$. This equation
can also be obtained by making the Ansatz $H = \sum_\nu E_\nu
B^\dagger_\nu B_\nu$ and demanding that $[H,B^\dagger_\nu] = E_\nu
B^\dagger_\nu$. For this auxiliary model the 
$B^\dagger_\nu$ thus describe independent single-particle excitations,
and the \qenergies\ $E_\nu$ are their eigenergies.

Let us now return to the Dicke model, where 
$X_\nu$ is an operator, so that we have to work a
little (but not much!) harder to satisfy \Eq{eq:condition1}.
To this end, commute $X_\nu$ past $P_{\nu+1}^n$ to the right and
use $X_\nu \vac = x_\nu \vac$, to obtain
\begin{subequations}
 \label{subeq:condition2}
\begin{eqnarray}
 \label{eq:condition2a}
\lefteqn{ \sum_{\nu=1}^n P_1^{\nu-1} 
P_{\nu+1}^n (\omega - E_\nu - 2 g x_\nu)  b^\dagger \vac }\\
& & =  
2 g  \sum_{\nu=1}^n P_1^{\nu-1}   [X_\nu, P_{\nu+1}^n]  b^\dagger \vac \; .
 \label{eq:condition2b}
\end{eqnarray}
\end{subequations}
To simplify the  second line, use \Eq{eq:commuteXpastP} and the
relation
\begin{eqnarray}
 \label{eq:XnuBmu}
 [X_\nu, B_\mu^\dagger] = - g \frac{B_\nu^\dagger - B_\mu^\dagger}{E_\nu - E_\mu} \; ,
\end{eqnarray}
which follows  from
$A_{\nu j} A_{\mu j} = -g (A_{\nu j} - A_{\mu j})/(E_\nu - E_\mu)$, 
to write $\sum_{\nu=1}^n P_1^{\nu-1}   [X_\nu, P_{\nu+1}^n]$ as 
\begin{subequations}
 \label{subeq:simplifyXnuPnu}
\begin{eqnarray}
 \label{eq:simplifyXnuPnu-b}
& &  - g 
\sum_{\nu=1}^n P_1^{\nu-1}  \sum_{\mu = \nu+1}^n  
P_{\nu+1}^{\mu-1} \frac{B_\nu^\dagger - B_\mu^\dagger}{E_\nu - E_\mu} 
P_{\mu+1}^n 
\\ 
 \label{eq:simplifyXnuPnu-c}
& & = \sum_{\nu=1}^n P_1^{\nu-1} P_{\nu+1}^n 
\sum_{\mu = 1, \mu \neq \nu}^n \frac{g}{E_\nu - E_\mu} \; . 
\end{eqnarray}
\end{subequations}
\Eq{eq:simplifyXnuPnu-c} follows by relabelling $\nu \leftrightarrow
\mu$ in the $B_\nu^\dagger$ term of \Eq{eq:simplifyXnuPnu-b}. 
Inserting  \Eq{eq:simplifyXnuPnu-c}
into \Eq{eq:condition2b}, we note that 
\Eq{subeq:condition2} is satisfied provided 
that the $n$ \qenergies\ $E_\nu$
obey the following $n$ coupled equations: 
\vspace{-3mm}
\begin{eqnarray}
\omega-E_{\nu} + \sum_{j=1}^N\frac{g^{2}}{E_\nu - \epsilon_{j}}
= \sum_{\mu=1,\mu\neq\nu}^{n}\frac{2g^{2}}{E_{\nu}-E_{\mu}} \; . \rule[-8mm]{0mm}{0mm}
\label{eq:BetheEqs}
\end{eqnarray}
These are the celebrated  Bethe-Ansatz equations for the 
Dicke model,  first obtained by Gaudin\cite{Gaudin}.
The fact that the right-hand side couples the equations for different
$E_\nu$ together presents the additional complication arising for the
Dicke model in comparison to the above-mentioned auxiliary boson
model. It implies that the $B^\dagger_\nu$ do \emph{not} describe 
independent single-particle excitations, since the value of any 
$E_\nu$ depends on that of all others. 

Generally \Eqs{eq:BetheEqs} have to be solved numerically. For sufficiently
small $n$, however, the original model
(\ref{eq:H}) can be diagonalised directly by solving the
eigenvalue problem in the basis of uncoupled bosonic and spin
eigenstates\cite{FewExcitationsDiagonalisation} instead of the basis
(\ref{eq:Ansatz}).

It is straightforward to expand the normalization factors of  Gaudin eigenstates\cite{Gaudin} and verify that 
$|\left\langle
   \Psi_{n}\middle|\Psi_{n}\right\rangle|^2  = \det\hat{M}$,
where $\hat M$ is an $n\times n$ matrix with elements
$M_{\nu \nu}  =  1+\sum_{j=1}^{N}A_{\nu j}^2 - 2 
\sum_{\mu = 1, \mu \neq \nu}^{n} A_{\mu \nu}^2$ and $M_{\mu \nu} = 2 A_{\mu \nu}^2$, 
and we used the shorthand $A_{\nu \mu} = g/(E_\nu - E_\mu)$.


We acknowledge support from Swiss NF, the NCCR Nanoscience Basel, and from the
DFG through SFB-TR12 and the cluster of excellence Nanosystems
Initiative Munich.  Part of this work was performed during the
workshop \textquotedblright{}From Femtoscience to Nanoscience: Nuclei,
Quantum Dots, and Nanostructures\textquotedblright{} in the Institute
of Nuclear Theory at the University of Washington.

\end{document}